\documentclass[12pt]{article}

\def\openone{\leavevmode\hbox{\small1\kern-3.8pt\normalsize1}}

\def\a{\alpha}
\def\b{\beta}

\def\d{\delta}
\def\e{\epsilon}

\def\g{\gamma}

\def\m{\mu}
\def\n{\nu}
\def\o{\omega}
\def\p{\pi}

\def\t{\tau}

\def\D{\Delta}

\def\G{\Gamma}

\def\L{\Lambda}

\def\ve{\varepsilon}

\def\cl{{\mathcal L}}

\def\bo{{\raise.15ex\hbox{\large$\Box$}}}               
\def\pr{\prod}                                          
\def\ltap{\raisebox{-.4ex}{\rlap{$\sim$}} \raisebox{.4ex}{$<$}}   
\def\face{{\raise.2ex\hbox{$\displaystyle \bigodot$}\mskip-2.2mu \llap {$\ddot
        \smile$}}}                                      
\def\dg{\dagger}                                     

\def\wt#1{\widetilde{#1}}                    
\def\Bar#1{\overline{#1}}                       
\def\VEV#1{\left\langle #1\right\rangle}        
\def\leftrightarrowfill{$\mathsurround=0pt \mathord\leftarrow \mkern-6mu
        \cleaders\hbox{$\mkern-2mu \mathord- \mkern-2mu$}\hfill
        \mkern-6mu \mathord\rightarrow$}       
\def\dvec#1{\vbox{\ialign{##\crcr
        \leftrightarrowfill\crcr\noalign{\kern-1pt\nointerlineskip}
        $\hfil\displaystyle{#1}\hfil$\crcr}}}           

\def\beq{\begin{equation}}
\def\eeq{\end{equation}}

\def\beqx{\begin{displaymath}}
\def\eeqx{\end{displaymath}}

\def\beqa{\begin{eqnarray}}
\def\eeqa{\end{eqnarray}}
\def\NO{\nonumber}

\def\pl#1#2#3{Phys.~Lett.~{\bf B {#1}}, #3 (19{#2})}
\def\np#1#2#3{Nucl.~Phys.~{\bf B {#1}}, #3 (19{#2})}
\def\prl#1#2#3{Phys.~Rev.~Lett.~{\bf #1}, #3 (19{#2})}
\def\pr#1#2#3{Phys.~Rev.~{\bf D {#1}}, #3 (19{#2})}
\newcommand{\prd}[2]{Phys.~Rev.~{\bf D#1}, #2 }

\def\prep#1#2#3{Phys.~Rep.~{\bf {#1}C}, #3 (19{#2})}

\def\ijmp#1#2#3{Int.~J.~Mod.~Phys.~{\bf A {#1}}, #3 (19{#2})}

\begin{document}
\date{\today}
\title{
\normalsize
\mbox{ }\hspace{\fill}
\begin{minipage}{4cm}
UPR-0870-T\\
{\tt hep-ph/0001204}
\end{minipage}\\[5ex]
{\large\bf Flavor Changing Effects in Theories with a 
Heavy $Z'$ Boson}\\[1ex] 
{\large\bf with Family Non-Universal Couplings}} 
\author{Paul Langacker and Michael Pl\"umacher\\
\em Department of Physics and Astronomy \\ 
\em University of Pennsylvania, Philadelphia PA 19104-6396, USA}
\date{}
\maketitle

\thispagestyle{empty}

\begin{abstract}
There are theoretical and phenomenological motivations that there may
exist additional heavy $Z'$ bosons with family non-universal
couplings. Flavor mixing in the quark and lepton sectors will then
lead to flavor changing couplings of the heavy $Z'$, and also of the
ordinary $Z$ when $Z-Z'$ mixing is included.  The general formalism of
such effects is described, and applications are made to a variety of
flavor changing and CP-violating tree and loop processes. Results are
described for three specific cases motivated by a specific heterotic
string model and by phenomenological considerations, including cases
in which all three families have different couplings, and those in
which the first two families, but not the third, have the same
couplings. Even within a specific theory the results are model
dependent because of unknown quark and lepton mixing matrices.
However, assuming that typical mixings are comparable to the CKM
matrix, processes such as coherent $\mu-e$ conversion in a muonic
atom, $K^0-\Bar{K}^{\,0}$ and $B-\Bar{B}$ mixing, $\epsilon$, and
$\epsilon'/\epsilon$ lead to significant constraints on $Z'$ bosons in
the theoretically and phenomenologically motivated range $M_{Z'} \sim$
1 TeV.
\end{abstract}

\clearpage

\section{Introduction}

Additional heavy neutral $Z'$ gauge bosons are amongst the best
motivated extensions of the standard model (SM), or its supersymmetric
extension (MSSM)~\cite{susyreview}. In particular, they often occur in
grand unified theories (GUTs), superstring theories, and theories with
large extra dimensions~\cite{xdimensions}.  In traditional GUTs, the
scale of the $Z'$ mass is arbitrary.  However, in perturbative
heterotic string models with supergravity mediated supersymmetry
breaking, the $U(1)'$ and electroweak breaking are both driven by a
radiative mechanism, with their scales set by the soft supersymmetry
breaking parameters, implying that the $Z'$ mass should be less than
around a TeV~\cite{TeVscale}. (The breaking can be at a larger
intermediate scale if it is associated with a $D$-flat
direction~\cite{intscale}.)  Furthermore, the extra symmetry can
forbid an elementary $\mu$ term, while allowing an effective $\mu$ and
$B \mu$ to be generated at the $U(1)'$ breaking scale, providing a
solution to the $\mu$ problem without introducing cosmological
problems~\cite{muprob,TeVscale}.  An extra $U(1)'$ provides an
analogous solution to the $\mu$ problem in models of gauge-mediated
supersymmetry breaking~\cite{gaugemed,cdm}.  An extra $U(1)'$ gauge
symmetry does not by itself spoil the successes of gauge coupling
unification.

There are stringent limits on the mass of an extra $Z'$ {}from the
non-observation of direct production followed by decays into $e^+ e^-$
or $\mu^+ \mu^-$ by CDF~\cite{CDF}, while indirect constraints {}from
precision data also limit the $Z'$ mass (weak neutral current
processes and LEP II) and severely constrain the $Z-Z'$ mixing angle
$\theta$ ($Z$-pole)~\cite{precision}.  These limits are
model-dependent, but are typically in the range $M_{Z'} > O(500) $ GeV
and $|\theta| <$ few $\times 10^{-3}$ for standard GUT models.
Recently, it has been argued~\cite{indications} that both $Z$-pole
data and atomic parity violation are much better described if the SM
or the MSSM is extended by an additional heavy $Z'$.

There is thus both theoretical and experimental motivation for an
additional $Z'$, most likely in the range 500 GeV - 1 TeV.  If true,
there should be a good chance to observe it at RUN II at the Tevatron,
and certainly at the LHC. Also, in this mass range, it should be
possible to carry out significant diagnostic probes of the $Z'$
couplings at the LHC and at a future NLC~\cite{diagnostics}, which
would complement those {}from the precision
experiments~\cite{indications}. The existence of a heavy $Z'$ would
also suggest a spectrum of sparticles considerably different than most
versions of the MSSM \cite{altspec}.

Most studies have assumed that the $Z'$ gauge couplings are family
universal~\cite{nonuniversal}, so that they remain diagonal
even in the presence of fermion flavor mixing by the GIM mechanism.
However, in string models it is possible to have family-nonuniversal
$Z'$ couplings, because of different constructions of the different
families. For instance, the consequences of a model by Chaudhuri,
Hockney, and Lykken~\cite{CHL} have been extensively analyzed
in~\cite{CHL5}, where it was shown that most $F$ and $D$-flat
directions involve an additional $U(1)'$ broken at the TeV scale. In
this case, the third generation quark couplings are different {}from the
first two families, and all three lepton generations have different
couplings.  Other aspects of the model are not realistic, but
nevertheless this provides a motivation to consider non-universal
couplings.  Similarly, possible anomalies in the $Z$-pole $b \bar{b}$
asymmetries~\cite{pdg} suggest that the data are better fitted with
a non-universal $Z'$~\cite{indications}.

Family-nonuniversal $Z'$ couplings necessarily lead to flavor-changing
(non-diagonal) $Z'$ couplings, and possibly to new CP-violating
effects, when quark and lepton flavor mixing are taken into
account\footnote{Models with an extra $Z'$ often include additional
exotic fermions (i.e., with non-standard $SU(2)$ assignments) as well.
Mixing of ordinary and exotic fermions could lead to flavor changing
$Z'$ and $Z$ couplings even in the absence of family non-universal
charges, and to non-universal $W$ couplings~\cite{exotic}. We do not
consider such effects in this paper.}. This will also imply flavor
violating $Z$ couplings if there is $Z-Z'$ mixing. Thus, flavor
changing neutral current (FCNC) and CP-violating effects should be
considered an additional constraint, consequence, or possibly
diagnostic probe of an extra $Z'$, and conversely as another
motivation to search for FCNC effects.

Even for a model in which the $Z'$ couplings are specified (prior to
flavor mixing), the predictions for FCNC are still model dependent,
because they depend on the individual unitary transformations for the
left ($L$) and right ($R$) chiral $u$-quarks, $d$-quarks, charged
leptons, and neutrinos which diagonalize their respective mass
matrices.  However, only the combination of $L$ matrices for the $u$
and $d$ occurring in the Cabibbo-Kobayashi-Maskawa (CKM) matrix is
known experimentally (with weak constraints on the leptonic analog
{}from neutrino oscillations), so one cannot make definitive
predictions. We will present our results for arbitrary mixings, and
then illustrate assuming that all of the mixing matrices are
comparable to the CKM matrix\footnote{For a complete theory the
$U(1)'$ charges would constrain the possible flavor mixings.  However,
such relations would be much more specific than the general issues
considered here.}.

In the next section we discuss the general formalism and introduce our
notation. In section 3 we discuss $Z'$ contributions to flavor
changing processes forbidden in the standard model and to new
contributions to SM processes, and we use experimental results to
constrain the $Z'$ couplings. Specific examples of a $Z'$ with flavor
changing couplings are discussed in section 4. Models in which all
three families have different $Z'$ charges in the gauge eigenstate
basis are strongly constrained by experimental results, and even
models in which the first two families have the same couplings, but
not the third, can yield flavor changing rates above experimental
limits unless they are suppressed by small mixing elements. Finally,
in section 5 we summarize our results and present our conclusions.

\section{Formalism}
We will use the formalism developed in ref.~\cite{Zprime} and
generalize it to the case of flavor violating $Z'$ couplings. In the
basis in which all the fields are gauge eigenstates the neutral current
Lagrangian is given by
\beq
  \cl_{NC}=-eJ^{\m}_{\mbox{\tiny em}}A_{\m} - g_1 J^{(1)\,\m} Z^{0}_{1,\m}
           - g_2 J^{(2)\,\m} Z^{0}_{2,\m}\;,
\eeq
where $Z^0_1$ is the $SU(2)\times U(1)$ neutral gauge boson,
$Z^0_2$ the new gauge boson associated with an additional Abelian
gauge symmetry, and the currents are
\beqa
  J^{(1)}_{\m}&=&\sum\limits_i \Bar{\psi}_i \g_{\mu} 
    \left[ \e_L(i)P_L + \e_R(i)P_R\right]\psi_i\;,\label{J1}\\[1ex]
  J^{(2)}_{\m}&=&\sum\limits_{i,j} \Bar{\psi}_i \g_{\mu} 
    \left[ \e^{(2)}_{{\psi_L}_{ij}}P_L 
     + \e^{(2)}_{{\psi_R}_{ij}}P_R\right]\psi_j\;,\label{J2}
\eeqa
where the sum extends over all quarks and leptons $\psi_{i,j}$
and $P_{R,L}=(1\pm\g_5)/2$. $\e^{(2)}_{{\psi_{R,L}}_{ij}}$ denote the
chiral couplings of the new gauge boson, and the standard model chiral
couplings are 
\beq
  \e_R(i)=-\sin^2\theta_WQ_i\;, \qquad
  \e_L(i)=t_3^i-\sin^2\theta_WQ_i\;,
\eeq
where $t_3^i$ and $Q_i$ are the third component of the weak isospin
and the electric charge of  fermion $i$, respectively. 
$g_1=g/\cos\theta_W=e/\sin\theta_W$ and $g_2$ are the gauge couplings
of the two $U(1)$ factors. 

Flavor changing effects (FCNCs) immediately arise if the $\e^{(2)}$
are non-diagonal matrices. If the $Z^0_2$ couplings are diagonal but
non-universal, flavor changing couplings are induced by fermion
mixing. The fermion Yukawa matrices $h_{\psi}$ in the weak eigenstate
basis can be diagonalized by unitary matrices $V^{\psi}_{R,L}$
\beq
  h_{\psi,diag}=V^{\psi}_R\,h_{\psi}\,{V^{\psi}_L}^{\dg}\;,
\eeq
where the 
CKM matrix is given by the combination
\beq
  V_{\rm CKM} = 
    V^u_L {V^d_L}^{\dg}\;.
  \label{BCKM}
\eeq
Hence, the chiral $Z^0_2$ couplings in the fermion mass eigenstate
basis read
\beq
  B^{\psi_L}_{ij}\equiv
    \left(V^{\psi}_L \e^{(2)}_{\psi_L} {V^{\psi}_L}^{\dg}\right)_{ij}\;,
    \qquad\mbox{and}\qquad
  B^{\psi_R}_{ij}\equiv
    \left(V^{\psi}_R \e^{(2)}_{\psi_R} {V^{\psi}_R}^{\dg}\right)_{ij}\;.
  \label{Bij}
\eeq

Further, $Z-Z'$ mixing is induced by electroweak symmetry breaking,
implying that $Z^0_{1,2}$ are related to mass eigenstates by an
orthogonal transformation. Hence, the couplings of the massive gauge
boson mass eigenstates $Z^{(i)}$ are
\beq
  \cl_{NC}^Z= - g_1 \left[\cos\theta J^{(1)\,\m} 
      + {g_2\over g_1}\sin\theta J^{(2)\,\m}\right] Z^{(1)}_{\m}
    - g_1\left[{g_2\over g_1}\cos\theta J^{(2)\,\m} 
      - \sin\theta J^{(1)\,\m}\right] Z^{(2)}_{\m}\;, \label{LZ}
\eeq
where $\theta$ is the $Z-Z'$ mixing angle.  The standard model weak
neutral current $J^{(1)\,\m}$ is given in eq.~(\ref{J1}), and
$J^{(2)\,\m}$ has a form analogous to eq.~(\ref{J2}), with the
$\e^{(2)}_{\psi_{R,L}}$ replaced by the couplings $B^{\psi_{R,L}}$
{}from eq.~(\ref{Bij}).

We have neglected kinetic mixing \cite{kinmix}, since it only amounts
to a redefinition of the unknown $Z'$ couplings.\footnote{Kinetic
mixing allows the redefined $Z^0_2$ charges to have a component of
weak hypercharge, which would otherwise not be allowed. This is
irrelevant for the purposes of this paper.}

At low energies, the effective four-fermion interactions are then given
by
\beqa
  -\cl_{eff}&=&{4G_F\over\sqrt{2}}\sum\limits_{\psi,\chi}
    \left(\rho_{eff} {J^{(1)}}^2 
    + 2wJ^{(1)}\cdot J^{(2)}+ y {J^{(2)}}^2\right)\\[1ex]
  &=&{4G_F\over\sqrt{2}}\sum\limits_{\psi,\chi}\sum\limits_{i,j,k,l}
  \left[
      C^{ij}_{kl} Q^{ij}_{kl}
    + \wt{C}^{ij}_{kl} \wt{Q}^{ij}_{kl}
    + D^{ij}_{kl} O^{ij}_{kl}
    + \wt{D}^{ij}_{kl} \wt{O}^{ij}_{kl}
  \right]\;,\label{Leff}
\eeqa
with the local operators\footnote{These operators are not all
independent. For couplings of four fermions of the same type,
$\psi=\chi$, e.g.\ four charged leptons, one has
$Q^{ij}_{kl}=Q^{kl}_{ij}$, $\wt{Q}^{ij}_{kl}=\wt{Q}^{kl}_{ij}$ and
$O^{ij}_{kl}=\wt{O}^{kl}_{ij}$.}
\beqa
  Q^{ij}_{kl}
    =\left(\Bar{\psi}_i\g^{\m}P_L\psi_j\right)
    \left(\Bar{\chi}_k\g_{\m}P_L\chi_l\right)\;,&\qquad&
  \wt{Q}^{ij}_{kl}
    =\left(\Bar{\psi}_i\g^{\m}P_R\psi_j\right)
    \left(\Bar{\chi}_k\g_{\m}P_R\chi_l\right)\;,\label{op}\\[1ex]
  O^{ij}_{kl}
    =\left(\Bar{\psi}_i\g^{\m}P_L\psi_j\right)
    \left(\Bar{\chi}_k\g_{\m}P_R\chi_l\right)\;,&\qquad&
  \wt{O}^{ij}_{kl}
    =\left(\Bar{\psi}_i\g^{\m}P_R\psi_j\right)
    \left(\Bar{\chi}_k\g_{\m}P_L\chi_l\right)\;.\NO
\eeqa
$\psi$ and $\chi$ represent classes of fermions with the same SM
quantum numbers, i.e., $u$, $d$, $e^-$, and $\nu$, while $i,j,k,l$ are
family indices. The coefficients are
\beqa
  C^{ij}_{kl}&=& \rho_{eff}\d_{ij}\d_{kl}\e_L(i)\e_L(k)
    + w\d_{ij}\e_L(\psi_i)B^{\chi_L}_{kl}
    + w\d_{kl}\e_L(\chi_l)B^{\psi_L}_{ij}
    + yB^{\psi_L}_{ij}B^{\chi_L}_{kl}\;,\\[1ex]
  \wt{C}^{ij}_{kl}&=& \rho_{eff}\d_{ij}\d_{kl}\e_R(i)\e_R(k)
    + w\d_{ij}\e_R(\psi_i)B^{\chi_R}_{kl}
    + w\d_{kl}\e_R(\chi_l)B^{\psi_R}_{ij}
    + yB^{\psi_R}_{ij}B^{\chi_R}_{kl}\;,\\[1ex]
  D^{ij}_{kl}&=& \rho_{eff}\d_{ij}\d_{kl}\e_L(i)\e_R(k)
    + w\d_{ij}\e_L(\psi_i)B^{\chi_R}_{kl}
    + w\d_{kl}\e_R(\chi_l)B^{\psi_L}_{ij}
    + yB^{\psi_L}_{ij}B^{\chi_R}_{kl}\;,\\[1ex]
  \wt{D}^{ij}_{kl}&=& \rho_{eff}\d_{ij}\d_{kl}\e_R(i)\e_L(k)
    + w\d_{ij}\e_R(\psi_i)B^{\chi_L}_{kl}
    + w\d_{kl}\e_L(\chi_l)B^{\psi_R}_{ij}
    + yB^{\psi_R}_{ij}B^{\chi_L}_{kl}\;.
\eeqa
The coefficients are given by
\beqa
  \rho_{eff}&=&\rho_1\cos^2\theta + \rho_2\sin^2\theta\;,
    \qquad \rho_i={M_W^2\over M_i^2\cos^2\theta_W}\;,\label{rho}\\[1ex]
  w&=&{g_2\over g_1}\sin\theta\cos\theta(\rho_1-\rho_2)\;,\label{w}\\[1ex]
  y&=&{\left(g_2\over g_1\right)}^2(\rho_1\sin^2\theta 
     + \rho_2\cos^2\theta)\;,\label{y}
\eeqa
where $M_i$ are the masses of the neutral gauge boson mass eigenstates
and $\theta_W$ is the electroweak mixing angle.

\section{Flavor Changing Processes}
In this section we will discuss flavor violating processes forbidden
in the SM and new contributions to SM processes. Experimental bounds
or results on these processes (cf.~ref.~\cite{pdg}) can then be used 
to constrain the $Z'$ couplings. 
\subsection{$Z$ Decays}
Due to the $Z-Z'$ mixing, $Z$ couples to $J^{(2)}$. The decay width
for a flavor changing $Z$ decay at tree level is given by
\beq
  \G(Z\to\psi_i\,\bar{\psi}_j)={CG_F\rho_1M_1^3\over3\sqrt{2}\p}
  \left({g_2\over g_1}\right)^2\sin^2\theta\left(
  \left|B^{\psi_L}_{ij}\right|^2 + \left|B^{\psi_R}_{ij}\right|^2\right)\;,
\eeq
where $C=1$ ($C=3$) is the color factor for leptons (quarks).  Due to
strong experimental constraints on the $Z-Z'$ mixing angle $\theta$,
cf.~ref.~\cite{precision,indications}, the $B^{\psi_{R,L}}$ cannot be
strongly constrained {}from flavor violating $Z$ decays.

\subsection{Lepton Decays}
In the SM each lepton generation has a separately conserved lepton
number, if one neglects small effects {}from non-vanishing neutrino
masses and non-perturbative effects. The effective Lagrangian
(\ref{Leff}) gives rise to lepton family number violating processes,
although the total lepton number is still conserved.

Consider first the decay of a charged lepton $l_j$ into three
different charged leptons $l_i$, $l_k$ and $\Bar{l}_l$. At tree level,
the decay width is
\beqa
  \G(l_j\to l_i\, l_k\, \Bar{l}_l)&=& {G_F^2 m_{l_j}^5\over 48 \p^3}\left(
  \left|C^{l_i\,l_j}_{l_k\,l_l}
       +C^{l_k\,l_j}_{l_i\,l_l}\right|^2 
  +\left|\wt{C}^{l_i\,l_j}_{l_k\,l_l}
       +\wt{C}^{l_k\,l_j}_{l_i\,l_l}\right|^2+\right.\NO\\[1ex]
  &&\left.+\left|D^{l_i\,l_j}_{l_k\,l_l}\right|^2 
  +\left|D^{l_k\,l_j}_{l_i\,l_l}\right|^2 
  +\left|\wt{D}^{l_i\,l_j}_{l_k\,l_l}\right|^2 
  +\left|\wt{D}^{l_k\,l_j}_{l_i\,l_l}\right|^2 \right)\;,
\eeqa
where we have neglected the masses of the final states leptons.
If two leptons in the final state are equal ($i=k$), taking
permutations of the external fermion lines into account yields
\cite{okada} 
\beq
\G(l_j\to l_i\, l_i\, \Bar{l}_l)= {G_F^2 m_{l_j}^5\over 48 \p^3}\left(
    2\left|C^{l_i\,l_j}_{l_i\,l_l}\right|^2 
  + 2\left|\wt{C}^{l_i\,l_j}_{l_i\,l_l}\right|^2
  + \left|D^{l_i\,l_j}_{l_i\,l_l}\right|^2 
  + \left|\wt{D}^{l_i\,l_j}_{l_i\,l_l}\right|^2\right)\;.
\eeq
Since such processes are free of hadronic uncertainties and well
constrained experimentally \cite{pdg,okada,tau}, they yield strong
constraints on the leptonic couplings of a $Z'$ \footnote{In
ref.~\cite{gan} these processes were considered in the case of
vanishing $Z$-$Z'$ mixing, $Z'$ couplings of the $V-A$ form, and
assuming that the $Z'$ has no diagonal couplings.}.

The strongest constraint on the $Z'$-$\mu$-$e$ coupling, however,
comes {}from coherent $\mu$-$e$ conversion in a muonic atom
\cite{sindrum}.  The branching fraction for this process, i.e., the
ratio of the coherent $\mu$-$e$ conversion rate to the $\mu$ capture
rate for a nucleus of atomic number $Z$ and neutron number $N$ is
given by
\cite{bernabeu,okada}
\beqa
  B(\mu^{-}N\to e^{-}N)&=&
    {G_F^2\a^3m_{\mu}^5\over2\p^2\G_{\mbox{\tiny CAPT}}}
    {Z_{eff}^4\over Z}\left|F_P\right|^2
    \left(
    \left|B^{l_L}_{12}\right|^2 + \left|B^{l_R}_{12}\right|^2
    \right)\Bigg|
    w\left[{1\over2}(Z-N)-2Z\sin^2\theta_W\right]+\NO\\[1ex]
  &&+y\left[
    (2Z+N)\left(B^{u_L}_{11}+B^{u_R}_{11}\right) + 
    (Z+2N)\left(B^{d_L}_{11}+B^{d_R}_{11}\right)
    \right]\Bigg|^2\;,
\eeqa
where $\G_{\mbox{\tiny CAPT}}$ is the $\mu$ capture rate, $Z_{eff}$ an
effective atomic charge obtained by averaging the muon wave function
over the nucleon, and $F_P$ is a nuclear matrix element.

\subsection{Radiative Decays}
Neutral current penguins give rise to radiative lepton decays.
Neglecting the mass of the final state lepton, the decay width is
\beq
  \G(l_j\to l_i\,\g)={\a G_F^2m_{l_j}^5\over8\p^4}\left(\,
    \left|\xi_L^{l_i\,l_j}\right|^2 + 
    \left|\xi_R^{l_i\,l_j}\right|^2 \,\right)\;,
\eeq
where the dipole moment couplings of an on-shell photon to the chiral
$\mu$-$e$ currents are 
\beqa
  \xi_L^{l_i\,l_j}&=&
      {1\over m_{l_j}}\sum\limits_k m_{l_k} D^{l_k\,l_j}_{l_i\,l_k}=
      {y\over m_{l_j}}\left(B^{l_R}m_lB^{l_L}\right)_{ij}
      +w\e_L(l_j)B^{l_R}_{ij}\;,\label{xiL}\\[1ex]
  \xi_R^{l_i\,l_j}&=&
      {1\over m_{l_j}}\sum\limits_k m_{l_k} \wt{D}^{l_k\,l_j}_{l_i\,l_k}=
      {y\over m_{l_j}}\left(B^{l_L}m_lB^{l_R}\right)_{ij}
      +w\e_R(l_j)B^{l_L}_{ij}\;,\label{xiR}
\eeqa
where $m_l$ is the charged lepton mass matrix.

A similar result holds for the decay $b\to s\g$. Since the
$b$-quark mass is much larger than the QCD-scale $\L$, long-range
strong interaction effects are not expected to be important in the
inclusive decay $B\to X_s\g$ \cite{buras_rev}. Hence, the rate for
this process is usually approximated by considering the ratio
\beq
  R\equiv {\G(B\to X_s\g)\over\G(B\to X_c\,e\,\Bar{\nu}_e\,)}
  \approx {\G(b\to s\g)\over\G(b\to c\,e\,\Bar{\nu}_e\,)}\;.
\eeq 
Neglecting SM contributions, the contribution to $R$ {}from the one-loop
neutral current penguin diagrams is
\beq
  R={8\a\over3\p}\,\left|V_{cb}\right|^{-2}\,
    f^{-1}\left({m_c^2\over m_b^2}\right)\,
    \left(\,\left|\xi_L^{s\,b}\right|^2 + 
    \left|\xi_R^{s\,b}\right|^2 \,\right)\;,
\eeq
where $f$ is the phase-space factor in the semi-leptonic $b$-decay:
\beq
  f(x) = 1-8x+8x^3-x^4-12x^2\ln x\;.
  \label{f}
\eeq
In analogy to eqs.~(\ref{xiL}) and (\ref{xiR}) the flavor violating
effective couplings $\xi_{R,L}^{s\,b}$ are given by:
\beqa
  \xi_L^{s\,b}&=&
      {1\over m_b}\sum\limits_k m_{d_k} D^{d_k\,b}_{s\,d_k}=
      {y\over m_b}\left(B^{d_R}m_dB^{d_L}\right)_{23}
      +w\e_L(b)B^{d_R}_{23}\;,\\[1ex]
  \xi_R^{s\,b}&=&
      {1\over m_b}\sum\limits_k m_{d_k} \wt{D}^{d_k\,b}_{s\,d_k}=
      {y\over m_b}\left(B^{d_L}m_dB^{d_R}\right)_{23}
      +w\e_R(b)B^{d_L}_{23}\;,
\eeqa
where $d_k$ stands for the $k$-th generation down-type quark and $m_d$
is the diagonal mass matrix of down quarks.

\subsection{Leptonic Meson Decays}
Meson decays can be used to place limits on the $Z'$ couplings to
quarks. Consider the lepton family number violating decay of a neutral
pseudoscalar meson $P^0$ into two charged leptons $l_i$ and
$\Bar{l_j}$, with $i\ne j$. Due to the hierarchy of lepton masses, we
can neglect the mass of the lighter lepton. Assuming that $m_{l_j}\ll
m_{l_i}$ (the case $m_{l_i}\ll m_{l_j}$ can be obtained by exchanging
the lepton indices $i$ and $j$ in the following) the decay width is
\beq
\G(P^0\to l_i\,\Bar{l}_j\,)=
    2{\G(P^-\to l_i\,\Bar{\n}_i\,)\over|V_{kl}|^2}
    \left(\left|\b_L^{l_i\,l_j}\right|^2
         +\left|\b_R^{l_i\,l_j}\right|^2\right)\;,
\eeq
where we have used isospin symmetry to relate the amplitude for
$P^0\to l_i\,\Bar{l}_j$ to the amplitude for the SM decay 
$P^-\to l_i\,\Bar{\n}_i$, and $V_{kl}$ is the element of the CKM-matrix
appearing in this SM process. The coefficients $\b_{R,L}$ for the
decays we have considered are given in table \ref{Ptable}.

In the SM the decay of a pseudoscalar $P^0$ into a lepton and its
anti-lepton is suppressed by the GIM mechanism and can only occur at
one-loop level (cf., e.g., ref.~\cite{Kdecays} for a discussion of
$K^0_L\to l_i\Bar{l_i}$ in the SM), whereas the $Z'$ couplings allow
tree-level contributions to such processes. Neglecting SM
contributions, which are formally of higher order in the couplings,
the decay width reads
\beq
  \G(P^0\to l_i\,\Bar{l_i})=
    4{\G(P^-\to l_i\,\Bar{\n_i})\over|V_{kl}|^2}\;
    {m_P^3\sqrt{m_P^2-4m_{l_i}^2}\over(m_P^2-m_{l_i}^2)^2}\;
    \left|\b_L^{l_i\,l_i}-\b_R^{l_i\,l_i}\right|^2\;,
\eeq
where the mass dependent factor corrects for the different phase
spaces in the decays of $P^0$ and $P^-$, and the couplings $\b_{R,L}$
can again be found in table \ref{Ptable}.

\begin{table}
\begin{center}
\begin{tabular}{c|c|c}
\hline
$P^0$ & $\b_L^{l_i\,l_j}$ & $\b_R^{l_i\,l_j}$\\[1ex]
\hline
&&\\[-1.5ex]
$\p^0$  & $D^{l_i\,l_j}_{u\,u}-C^{l_i\,l_j}_{u\,u}
          -D^{l_i\,l_j}_{d\,d}+C^{l_i\,l_j}_{d\,d}$ 
        & $\wt{C}^{l_i\,l_j}_{u\,u}-\wt{D}^{l_i\,l_j}_{u\,u}
          -\wt{C}^{l_i\,l_j}_{d\,d}+\wt{D}^{l_i\,l_j}_{d\,d}$\\[2ex]
$K^0_L$ & $D^{l_i\,l_j}_{d\,s}+D^{l_i\,l_j}_{s\,d}
          -C^{l_i\,l_j}_{d\,s}-C^{l_i\,l_j}_{s\,d}$
        & $-\wt{D}^{l_i\,l_j}_{d\,s}-\wt{D}^{l_i\,l_j}_{s\,d}
          +\wt{C}^{l_i\,l_j}_{d\,s}+\wt{C}^{l_i\,l_j}_{s\,d}$\\[2ex]
$D^0$   & $\sqrt{2}\left(D^{l_i\,l_j}_{c\,u}-C^{l_i\,l_j}_{c\,u}\right)$
        & $\sqrt{2}\left(-\wt{D}^{l_i\,l_j}_{c\,u}
          +\wt{C}^{l_i\,l_j}_{c\,u}\right)$ \\[2ex]
$B^0$   & $\sqrt{2}\left(D^{l_i\,l_j}_{b\,d}-C^{l_i\,l_j}_{b\,d}\right)$
        & $\sqrt{2}\left(-\wt{D}^{l_i\,l_j}_{b\,d}
          +\wt{C}^{l_i\,l_j}_{b\,d}\right)$ \\[2ex]
$B^0_s$ & $\sqrt{2}\left(D^{l_i\,l_j}_{b\,s}-C^{l_i\,l_j}_{b\,s}\right)$
        & $\sqrt{2}\left(-\wt{D}^{l_i\,l_j}_{b\,s}
          +\wt{C}^{l_i\,l_j}_{b\,s}\right)$\\[2ex]
\hline
\end{tabular}
\end{center}
\caption{Coefficients in the decay widths of pseudoscalar mesons $P^0$. 
  Since $K^0_L$ is a linear combination of $K^0$ and $\Bar{K}^{\,0}$, the 
  $\b_{R,L}$ for $K^0_L$ decays depend only on the real part of the
  $Z'$-$d$-$s$ axial vector coupling Re$(B_{12}^{d_R}-B_{12}^{d_L})$.
  \label{Ptable}} 
\end{table}

Similar formulae hold for semi-leptonic $\t$ decays. For the
process $\t\to l_i\, \p^0$ we have
\beq
  \G(\t\to l_i\, \p^0)=2{\G(\t\to\n_{\t}\p^{-})\over |V_{ud}|^2}
    \left(\left|\b_L^{l_i\,\t}\right|^2
         +\left|\b_R^{l_i\,\t}\right|^2\right)\;,
\eeq
where we have neglected the mass of the final state lepton and the 
$\b_{R,L}$ are given in the first line of table \ref{Ptable},
whereas for $\t\to l_i\, K^0$ one finds
\beq
  \G(\t\to l_i\, K^0)=4{\G(\t\to\n_{\t}K^{-})\over |V_{us}|^2}
    \left(\left|
    C^{l_i\,\t}_{d\,s}-D^{l_i\,\t}_{d\,s}
    \right|^2+\left|
    \wt{C}^{l_i\,\t}_{d\,s}-\wt{D}^{l_i\,\t}_{d\,s}
    \right|^2\right)\;.
\eeq
Replacing the indices $d$ and $s$ yields the decay width for 
$\t\to l_i\,\Bar{K}^{\,0}$. 

\subsection{Semi-Leptonic Meson Decays}
All the processes discussed in the last section constrain only
couplings of the form $D^{l_i\,l_j}_{q_k\,q_l}-
C^{l_i\,l_j}_{q_k\,q_l}$, i.e.,\ the axial vector couplings in the
quark current. Limits on the corresponding vector couplings can be
obtained by considering decays of $P^0$ into another pseudoscalar
meson and two leptons.

\begin{table}
\begin{center}
\begin{tabular}{c|c|c}
\hline
$P$ & $\d_L^{l_i\,l_j}$ & $\d_R^{l_i\,l_j}$\\[1ex]
\hline
&&\\[-1.5ex]
$K^0_L$ & $D^{l_i\,l_j}_{d\,s}-D^{l_i\,l_j}_{s\,d}
          +C^{l_i\,l_j}_{d\,s}-C^{l_i\,l_j}_{s\,d}$
        & $\wt{D}^{l_i\,l_j}_{s\,d}-\wt{D}^{l_i\,l_j}_{d\,s}
          +\wt{C}^{l_i\,l_j}_{s\,d}-\wt{C}^{l_i\,l_j}_{d\,s}$\\[2ex]
$D^0$   & $\sqrt{2}\left(D^{l_i\,l_j}_{c\,u}+C^{l_i\,l_j}_{c\,u}\right)$
        & $\sqrt{2}\left(\wt{D}^{l_i\,l_j}_{c\,u}
          +\wt{C}^{l_i\,l_j}_{c\,u}\right)$ \\[2ex]
$B^0$   & $\sqrt{2}\left(D^{l_i\,l_j}_{b\,s}+C^{l_i\,l_j}_{b\,s}\right)$
        & $\sqrt{2}\left(\wt{D}^{l_i\,l_j}_{b\,s}
          +\wt{C}^{l_i\,l_j}_{b\,s}\right)$ \\[2ex]
\hline
\end{tabular}
\end{center}
\caption{Coefficients for semi-leptonic meson decays. The 
  $\d_{R,L}$ for $K^0_L$ decays are proportional to the imaginary 
  part of the $Z'$-$d$-$s$ vector coupling 
  Im$(B_{12}^{d_R}+B_{12}^{d_L})$.\label{Ptable2}} 
\end{table}

Particularly interesting are lepton flavor conserving, $CP$ violating
contributions to decays $K_L^0\to\pi^0\bar{l}l$, since the branching
ratios are expected to be small in the SM \cite{Kdecays,Ktopiee} and
new limits {}from KTeV \cite{KTEV} allow the imaginary part of the
$Z'$-$d$-$s$ vector coupling to be constrained.  Neglecting the
electron mass but taking the $\mu$-mass into account, the decay widths
for the semi-leptonic $K_L^0$ decays considered are
\beqa
  \G(K_L^0\to e^+e^-\p^0\,)&=&2{\G(K^+\to e^+\nu_e\p^0\,)\over|V_{us}|^2}
    \left(|\d_L^{e\,e}|^2+|\d_R^{e\,e}|^2\right)\;,\\[1ex]
  \G(K_L^0\to \mu^+\mu^-\p^0\,)&=&2{\G(K^+\to\mu^+\nu_{\mu}\p^0\,)
    \over|V_{us}|^2}
    \left[0.57\left(|\d_L^{\mu\,\mu}|^2+|\d_R^{\mu\,\mu}|^2\right)
      -0.48\,\mbox{Re}(\d_L^{\mu\,\mu}{\d_R^{\mu\,\mu}}^*)\right]\;,
\eeqa
where the numerical coefficients in the last decay width arise due to
the different phase spaces for the processes $K_L^0\to \mu^+\mu^-\p^0$
and $K^+\to\mu^+\nu_{\mu}\p^0$, and the couplings $\d_{R,L}$ for these
processes are given in table \ref{Ptable2}. 
We also considered the lepton flavor violating decay
\beq
  \G(K_L^0\to \mu^+e^-\p^0\,)=2{\G(K^+\to \mu^+\nu_{\mu}\p^0\,)
    \over|V_{us}|^2}\left(|\d_L^{e\,\mu}|^2+|\d_R^{e\,\mu}|^2\right)\;,
\eeq
although experimental bounds on the $Z'$-$\mu$-$e$ coupling {}from coherent 
$\mu$-$e$ conversion imply that the branching ratio for this process is
several orders of magnitude below the experimental bounds.

Similarly, for semi-leptonic $D^0$ and $B^0$ decays one has
\beqa
  \G(D^0\to l_i\Bar{l}_j\p^0)&=&{\G(D^+\to\Bar{l}_j\nu_j\p^0)\over|V_{cd}|^2}
    \left(|\d_L^{l_i\,l_j}|^2+|\d_R^{l_i\,l_j}|^2\right)\;,\\[1ex]
  \G(B^0\to l_i\Bar{l}_jK^0)&=&{\G(B^+\to\Bar{l}_i\nu_i\Bar{D}^{\,0})
    \over|V_{cb}|^2}\,{f(m_K^2/m_B^2)\over f(m_D^2/m_B^2)}
    \left(|\d_L^{l_i\,l_j}|^2+|\d_R^{l_i\,l_j}|^2\right)\;,
\eeqa
where the phase-space function $f$ is given in eq.~(\ref{f}).

\subsection{Mass Splittings and CP Violation}
The effective Lagrangian (\ref{Leff}) also contributes to the mass
splitting in a neutral pseudo-scalar meson system. Again denoting the
flavor eigenstates of a meson by $P^0$ and $\Bar{P}^{\,0}$, the mass
splitting $\D m_P$ is given by
\beq
  \D m_P=-2\mbox{Re}\langle P^0 | \cl_{eff} |\,\Bar{P}^{\,0}\,\rangle\;.
\eeq
The relevant hadronic matrix elements of the operators (\ref{op}) have
been determined in the vacuum insertion approximation using PCAC
\cite{DM_matr_elem}. Hence, for a meson with the quark content
$P^0=\Bar{q}_jq_i$ we obtain the following contribution to the mass
splitting:
\beq
  \D m_P=4\sqrt{2}G_Fm_PF_P^2y\left\{
  {1\over3}\mbox{Re}\left[\left(B^{q_L}_{ij}\right)^2
    +\left(B^{q_R}_{ij}\right)^2\right]-
  \left[{1\over2}+
    {1\over3}\left({m_P\over m_{q_i}+m_{q_j}}\right)^2\right]
  \mbox{Re}\left(B^{q_L}_{ij}B^{q_R}_{ij}\right)\right\}\;,
\eeq
where $m_P$ and $F_P$ are the mass and decay constant of the meson,
respectively.

Further, phases in the $Z'$ couplings $B^{\psi_{R,L}}_{ij}$ will
contribute to CP violating processes. Limits on the imaginary parts of
the $s$-$d$-$Z'$ couplings can be placed by considering indirect CP
violation $\ve_K$ in the neutral kaon system:
\beqa
  |\ve_K|&=&{1\over2\sqrt{2}}{
    \mbox{Im}\langle K^0 | \cl_{eff} |\,\Bar{K}^{\,0}\,\rangle 
    \over
    \mbox{Re}\langle K^0 | \cl_{eff} |\,\Bar{K}^{\,0}\,\rangle}\\[1ex]
  &=&{2G_Fm_KF_K^2y\over\D m_K}\left|
  {1\over3}\mbox{Im}\left[\left(B^{d_L}_{12}\right)^2
    +\left(B^{d_R}_{12}\right)^2\right]-
  \left[{1\over2}+
    {1\over3}\left({m_P\over m_d+m_s}\right)^2\right]
  \mbox{Im}\left(B^{d_L}_{12}B^{d_R}_{12}\right)
  \right|\;.
\eeqa
Direct CP violation $\ve'$ in the decays $K\to\p\p$ can be expressed
in terms of the decay amplitudes $A_0=A(K\to(\p\p)_0)$ and
$A_2=A(K\to(\p\p)_2)$, where the indices $0$ and $2$ denote the
isospin of the final two pion state (see~ref.~\cite{buras_rev} for a 
review):
\beq
  \ve'=-{1\over\sqrt{2}}{\o\over\mbox{Re}A_0}\left(
      \mbox{Im}A_0-{1\over\o}\mbox{Im}A_2\right)
      \mbox{e}^{i\tilde{\phi}}\;,\label{eprime}
\eeq
where
\beq
  \o={\mbox{Re}A_2\over\mbox{Re}A_0}\;,\qquad 
  \tilde{\phi}={\p\over2}+\d_2-\d_0\;.
\eeq
The $\d_I$ are the final state interaction phases. When using
eq.~(\ref{eprime}) to constrain physics beyond the standard model, it
is common practice to take $\o$, $\mbox{Re}A_0$ and $\tilde{\phi}$
{}from experiment,
\beq
  \o=0.045\;,\qquad\mbox{Re}A_0=3.33\cdot10^{-7}\,\mbox{GeV}\;,
  \qquad\tilde{\phi}\approx{\p\over4}\;,
\eeq
and consider new contributions to the imaginary parts of the
amplitudes $A_0$ and $A_2$. This is due to the fact that the CP
violating imaginary parts are dominated by short-distance effects and
can be reliably determined by considering matrix elements of the
effective Lagrangian (\ref{Leff}). The hadronic matrix elements can be
computed in the large $N_c$ limit of chiral perturbation theory
(cf.~appendix \ref{KTOPI}), and one finds the following neutral current
contribution to the real part of the ratio $\ve'/\ve_K$:
\beqa
  &&\mbox{Re}\left({\ve'\over\ve_K}\right)=
    2\cdot10^3\,w\left(\mbox{Im}B^{d_L}_{21}
    +{3\over2}\mbox{Im}B^{d_R}_{21}\right)+\\[1ex]
  &&+1.5\cdot10^3\,y\left[\left(B^{u_L}_{11}-B^{d_L}_{11}\right)\left(
      \mbox{Im}B^{d_L}_{21}+2\mbox{Im}B^{d_R}_{21}\right)
    -\left(B^{u_R}_{11}-B^{d_R}_{11}\right)\left(
      2\mbox{Im}B^{d_L}_{21}+\mbox{Im}B^{d_R}_{21}\right)\right]\;.\NO
\eeqa

\subsection{Experimental Constraints \protect\label{exp}}
Experimental limits or results on these processes can be used to
constrain the flavor violating $Z'$ couplings\footnote{Flavor diagonal
$Z'$ couplings can be constrained {}from fits to electroweak
observables \cite{precision,indications}.}. In the following we
briefly discuss bounds coming {}from $Z$-$Z'$ mixing contributions to
these processes. The pure $Z'$ contributions yield a multitude of
bounds on products of $Z'$ couplings which are less illuminating. In
the examples that we discuss in the next section, these contributions
are of the same order as the mixing contributions.

As already mentioned, the strongest bound on the $Z'$-$\mu$-$e$
coupling comes {}from the non-observation of coherent $\mu$-$e$
conversion by the Sindrum-II collaboration \cite{sindrum}:
\beq
  w^2\left(\left|B^{l_L}_{12}\right|^2+\left|B^{l_R}_{12}\right|^2
    \right)<4\cdot10^{-14}\;,\label{mutoe}
\eeq 
while the decays $\t\to3e$ and $\t\to3\mu$ yield the strongest bounds
on flavor violating $\t$ couplings:
\beq
  w^2\left(\left|B^{l_L}_{13}\right|^2+\left|B^{l_R}_{13}\right|^2
    \right)<2\cdot10^{-5}\;,\qquad
  w^2\left(\left|B^{l_L}_{23}\right|^2+\left|B^{l_R}_{23}\right|^2
    \right)<10^{-5}\;.
\eeq
It is interesting to note that these constraints alone ensure that
branching ratios for lepton flavor violating meson decays are below
the experimental bounds, provided that the
parameters $w$ and $y$, given in eqs.~(\ref{w}) and (\ref{y}), are of
the same order. (This holds in the most interesting case of a TeV
scale $Z'$ with small mixing, $\theta\ltap10^{-3}$.)
For example, upper limits on the branching ratios
for the processes $K_L\to\mu^{\pm}e^{\mp}$ {}from the BNL E871
collaboration \cite{bnl} and $K_L\to\pi^0\mu^{\pm}e^{\mp}$ {}from KTeV 
\cite{buchalla} yield
\beqa
  y^2\left(\left|B^{l_L}_{12}\right|^2
    +\left|B^{l_R}_{12}\right|^2\right)
    \left|\mbox{Re}B^{d_R}_{12}
    -\mbox{Re}B^{d_L}_{12}\right|^2&<&10^{-14}\;,\label{k1}\\[1ex]
  y^2\left(\left|B^{l_L}_{12}\right|^2
    +\left|B^{l_R}_{12}\right|^2\right)
    \left|\mbox{Im}B^{d_R}_{12}
    +\mbox{Im}B^{d_L}_{12}\right|^2&<&2\cdot10^{-10}\;.\label{k2}
\eeqa
Hence, the experimental bounds on these processes would have to be
improved by several orders of magnitude to yield interesting
constraints on the real and imaginary parts of $B^{d_{R,L}}_{12}$.
{}From eqs.~(\ref{mutoe}), (\ref{k1}) and (\ref{k2}) it is clear that
lepton flavor violating meson decays cannot compete in constraining
flavor non-diagonal $Z'$ couplings, except in the limit
$|w| \ll y$.

However, lepton flavor conserving meson decays can be used to
constrain the $Z'$ couplings to quarks, e.g., limits on
$K_L\to\mu^+\mu^-$ \cite{pdg} and $K_L\to\pi^0\mu^+\mu^-$ \cite{KTEV}
give:
\beq
  w^2\left|\mbox{Re}B^{d_R}_{12}
    -\mbox{Re}B^{d_L}_{12}\right|<3\cdot10^{-11}\;,\qquad
  w^2\left|\mbox{Im}B^{d_R}_{12}
    +\mbox{Im}B^{d_L}_{12}\right|^2<5\cdot10^{-11}\;.
\eeq

The most stringent bounds on the absolute values of the remaining
non-diagonal $Z'$ couplings to quarks then come {}from decays of $D^0$
and $B^0$ into a $\mu^+\mu^-$ pair \cite{pdg} and from the process
$B^0\to K^0\mu^+\mu^-$ \cite{cleo}:
\beq
  w^2\left|B^{u_{R,L}}_{12}\right|^2<6\cdot10^{-4}\;,\qquad
  w^2\left|B^{d_{R,L}}_{13}\right|^2<10^{-5}\;,\qquad
  w^2\left|B^{d_{R,L}}_{23}\right|^2<3\cdot10^{-6}\;.
\eeq
The top-quark couplings to a $Z'$ cannot be constrained {}from these
tree-level processes. In the future, studies of rare top decays
\cite{topdecays} and associated top-charm production \cite{topcharm} 
at the Tevatron, LHC and a future $e^+e^-$ linear collider will yield
very useful constraints.

Further, experimental results on meson mass splittings\footnote{The
$B_s$-$\Bar{B}_s$ mass difference has not been measured yet. We
required that new contributions be smaller than the lower limit 
on the mass splitting, $\D m_{B_s^0}> 14.3\,\mbox{ps}^{-1}$ at 
$95 \%$ C.L.\ \cite{Bsmixing}.} allow constraints on the 
real parts of the squared $Z'$ couplings to quarks,
\beqa
  y\,\left|\mbox{Re}\left[\left(B^{d_{R,L}}_{12}\right)^2\right]\right|
    <10^{-8}\;,&\qquad&
  y\,\left|\mbox{Re}\left[\left(B^{d_{R,L}}_{13}\right)^2\right]\right|
    <6\cdot10^{-8}\;,\\[1ex]
  y\,\left|\mbox{Re}\left[\left(B^{d_{R,L}}_{23}\right)^2\right]\right|
    <2\cdot10^{-6}\;,&\qquad&
  y\,\left|\mbox{Re}\left[\left(B^{u_{R,L}}_{12}\right)^2\right]\right|
    <10^{-7}\;,
\eeqa
and $CP$ violation in the Kaon system yields constraints on the
imaginary part of the $Z'$-$d$-$s$ coupling:
\beq
  y\,\left|\mbox{Im}\left[\left(B^{d_{R,L}}_{12}\right)^2\right]\right|
    <8\cdot10^{-11}\;,\qquad
  w\,\left|\mbox{Im}B^{d_{R,L}}_{12}\right|<10^{-6}\;.
\eeq

\section{Models}
In the following we shall study concrete examples of extended Abelian
gauge structures with flavor non-universal couplings, in order to see
where such effects are most likely to be seen. Although we only
discussed bounds coming {}from $Z$-$Z'$ mixing contributions to FCNC
processes in section \ref{exp}, we will also take pure $Z'$
contributions into account here, since they are of the same order as
mixing contributions in the models considered.

First we consider a perturbative heterotic superstring model, based on
the free fermionic construction \cite{CHL}. Such models have been
studied in detail \cite{CHL5} and it was shown that they generically
contain extended Abelian gauge structures and additional matter at the
string scale. The running of a scalar mass-square due to large Yukawa
couplings then triggers the radiative breaking of the $U(1)'$,
naturally giving a $Z'$ in the TeV mass range.

The $Z'$ couplings can be calculated and the fermion charges $Q'$ can
be found in table \ref{chl5_table}. In the quark sector, the first two
generations have the same charges, i.e., in the fermion mass
eigenstate basis only mixings of the third generation quarks induce
flavor changing quark-couplings in eq.~(\ref{Bij}).  Nevertheless, all
the $B^q_{ij}$ are nonzero in general.  The same holds true for
right-handed leptons, but all three left-handed lepton generations
have different $Q'$ charges, which could give rise to strong flavor
violating effects.

To study these FCNCs we have chosen a $Z'$ mass of $1\;$TeV and a
$Z$-$Z'$ mixing angle $\theta=10^{-3}$. The $Z'$ coupling strength,
predicted {}from the string model, is $g_2=0.105$ \cite{CHL5}.
Further, we have to specify the unknown fermion mixing matrices
$V^{\psi}_{R,L}$. As an example, we will assume that they are equal to
the CKM matrix.

In the charged lepton sector these couplings then predict rates for
flavor violating processes which are six orders of magnitude above the
experimental limits for coherent $\mu$-$e$ conversion, five orders of
magnitude above the limit for the decay $\mu\to3e$, and of the same
order as the recent experimental bound {}from the MEGA collaboration
\cite{MEGA} for the radiative decay $\mu\to e\g$. On the other hand,
predictions for flavor violating $\t$ decays are well below the
experimental limits. This is due in part  to the fact that these bounds
are much less restrictive than for the muon, and in part to the
assumed CKM mixing, where the 13 and 23 elements are rather small.
Assuming larger mixing of third generation leptons, as suggested by
the atmospheric neutrino data \cite{atmo}, would give flavor
changing rates close to the experimental bounds, particularly for $\t$
decays into three charged leptons.

For processes involving quarks, we obtain contributions of the same
order as SM contributions for the $B$-$\Bar{B}$ and $B_s$-$\Bar{B}_s$
mass differences, and, assuming maximal $CP$ violation, a contribution
to $\ve_K$ which is of the same order as the measured value.
Predictions for lepton flavor violating meson decays are well below
the experimental bounds.

\begin{table}[t]
\begin{center}
\begin{tabular}{|c|r|}
  \multicolumn{2}{c}{string model}\\
  \hline
  multiplet & $100\, Q^\prime$ \\ \hline
  $\left( \begin{array}{c} t \\ b \end{array} \right)_L$ & $- 71$ \\[1ex]
  $t_R$                                                  & $+133$ \\[1ex]
  $b_R$                                                  & $-136$ \\[1ex]
  $\left( \begin{array}{c} u \\ d \end{array} \right)_L$,
  $\left( \begin{array}{c} c \\ s \end{array} \right)_L$ & $+ 68$ \\[1ex]
  $u_R$, $c_R$                                           & $-  6$ \\[1ex]
  $d_R$, $s_R$                                           & $+  3$ \\[1ex]
  $\left(\begin{array}{c}\nu_\tau\\\tau\end{array}\right)_L$&$+74$\\[1ex]
  $\tau_R$                                              & $-130$ \\[1ex]
  $\left(\begin{array}{c}\nu_\mu\\\mu\end{array}\right)_L$&$- 65$\\[1ex]
  $\mu_R$                                               & $+  9$ \\[1ex]
  $\left(\begin{array}{c}\nu_e\\ e\end{array}\right)_L$ & $-204$ \\[1ex]
  $e_R$                                                 & $+  9$\\[1ex]
  \hline
\end{tabular}
\hspace{2cm}
\begin{tabular}{|c|r|}
  \multicolumn{2}{c}{EW fit model}\\
  \hline
  multiplet & $100\,Q^\prime$ \\ \hline
  $\left( \begin{array}{c} t \\ b \end{array} \right)_L$ & $+132$ \\[1ex]
  $t_R$                                                  & $+100$ \\[1ex]
  $b_R$                                                  & $+848$ \\[1ex]
  $\left( \begin{array}{c} u \\ d \end{array} \right)_L$,
  $\left( \begin{array}{c} c \\ s \end{array} \right)_L$ & $-52$ \\[1ex]
  $u_R$, $c_R$                                           & $+38$ \\[1ex]
  $d_R$, $s_R$                                           & $+172$ \\[1ex]
  $\left(\begin{array}{c}\nu_\tau\\\tau\end{array}\right)_L$&$-24$\\[1ex]
  $\tau_R$                                              & $+3$ \\[1ex]
  $\left(\begin{array}{c}\nu_\mu\\\mu\end{array}\right)_L$&$- 32$\\[1ex]
  $\mu_R$                                               & $-31$ \\[1ex]
  $\left(\begin{array}{c}\nu_e\\ e\end{array}\right)_L$ & $-32$ \\[1ex]
  $e_R$                                                 & $-31$\\[1ex]
  \hline
\end{tabular}
\end{center}
\caption{Fermion charges in the $Z'$ models motivated {}from string
  theory and {}from precision electroweak data. 
  \label{chl5_table}}
\end{table}

As we have seen, one obtains flavor violating rates above the
experimental limits in the lepton sector if the first two generations
have different $Q'$ charges. As an example of a model in which the
first two quark families also have different charges, we again
consider the string motivated model; however, we set the charges to
zero by hand for all of the first generation fermions. Then the rates
for coherent $\mu$-$e$ conversion and $\mu\to3e$ are still too large
by four and two orders of magnitude, respectively, and we find the
same contributions as before to the $B$-$\Bar{B}$ and
$B_s$-$\Bar{B}_s$ mass differences. In addition, however, we obtain
contributions to the mass splitting in the $K$- and $D$-systems, which
are larger than the measured values by two orders of magnitude, and
the predicted rates for lepton flavor violating and conserving $K_L$
decays are well above experimental limits or results. Further, again
assuming maximal $CP$ violation, we have contributions to $\ve_K$ and
Re$(\ve'/\ve_K)$ which are too large by factors $6\cdot10^{5}$ and
$20$, respectively.

{}From these examples, we conclude that any TeV-scale $Z'$ would
almost certainly have to have equal couplings to the first two
families. However, there is still the possibility of different
couplings for the third family.

As a final example we consider a flavor non-universal $Z'$ that was
recently shown to improve the fit to precision electroweak data
\cite{indications}. Assuming that the first two fermion generations
are flavor universal\footnote{Such models arise for example in the
framework of $E_6$ models, if discrete symmetries which lead to small
neutrino masses are imposed \cite{E6nardi}.}, one can determine the
$Z'$ couplings {}from the fit. The central values found in
ref.~\cite{indications} are reproduced in table
\ref{chl5_table}. Since the $Z'$ coupling of right-handed top-quarks
was not determined we have set $Q'_{t_R}=1$ for definiteness, although
this coupling has only very little influence on the processes we
discussed.

Since in this model the first two lepton generations have the same
$Z'$ couplings, the predicted rates for flavor violating $\mu$ decays
are well below the experimental limits. Only for coherent $\mu$-$e$
conversion do we find a predicted rate of the same order as the
experimental limit. However, we obtain contributions to the
$B$-$\Bar{B}$ and $B_s$-$\Bar{B}_s$ mass differences which are too
large by factors $7$ and $40$, respectively. Further, the predicted
value for $\ve_K$ is larger than the measured value by a factor $20$
and there is a contribution to $\ve'$ which is of the same order as
the measured one.  Finally, the branching ratios predicted for lepton
flavor conserving decays $K_L\to\pi^0l^+l^-$ and $B^0\to K^0l^+l^-$
are only two orders of magnitude below the experimental bounds, i.e.,
further experimental progress on these processes could help to
constrain this model.

Thus, even with universal couplings for the first two families, mixing
with the third family induces significant effects in the first two
families, at least if the fermion mixing matrices for the charged
leptons and the $d$-type quarks are comparable to the CKM matrix.

\section{Conclusions}
We conclude that additional $Z'$ bosons with a TeV scale mass and
family non-universal couplings are severely constrained by
experimental results on flavor changing processes. The most stringent
bounds come {}from muon decays, coherent $\mu-e$ conversion in muonic
atoms, and {}from lepton flavor conserving processes in the neutral
$K$-system, i.e.,\ {}from processes involving the coupling of a $Z'$
to first and second generation fermions. Couplings to the third
generation are less constrained, but future studies of rare top,
bottom and $\tau$ decays will help to further constrain these models.

If the $Z'$ couplings are diagonal but family non-universal in the
gauge eigenstate basis, flavor changing couplings arise due to fermion
mixing. In the examples we assumed that these unknown mixing matrices
are comparable to the CKM matrix. If all three families have different
couplings we find contributions to flavor changing processes involving
the first two generations which are above experimental bounds by
several orders of magnitude. We obtain particularly large
contributions to coherent $\mu$-$e$ conversion, the decay $\mu\to3e$,
meson mass splittings, $K_L$ decays and $CP$ violation in the neutral
$K$-system.

Since couplings of third generation fermions are much less constrained
and we assumed that the fermion mixing matrices have a structure
similar to the CKM matrix, these problems can be alleviated by
assuming that the first two families, but not the third, have the same
$U(1)'$ charges.  Mixing with the third family still induces flavor
changing effects involving the first two families, but they are
suppressed since in the CKM matrix those mixings are small.  In the
model considered, the new contributions are too large for the
$B$-$\Bar{B}$ and $B_s$-$\Bar{B}_s$ mass differences, and
$CP$-violation in the neutral $K$-system.  The experimental bounds for
all other processes are respected.

All of the constraints are model dependent. In addition to the $Z'$
mass, mixing with the $Z$, and charges, they are dependent on the
mixing matrices for the left and right chiral quarks and leptons.  In
the standard model, the right chiral mixing matrices are unobservable,
and only the combinations of left chiral matrices in the CKM matrix
(\ref{BCKM}) and its leptonic analog are observable.  However, all of
these matrices are in principle observable in the presence of
non-universal $Z'$ couplings.  For example, the flavor changing
effects in the $B$ and $K$ systems could be eliminated if the CKM
mixing were due entirely to the $u$ quark sector, i.e., 
$ V_{\rm CKM} = V^u_L$, with $ V^d_L = V^d_R = \openone$. Similarly,
$\mu$-$e$ conversion and the decay $\mu\to3e$ would be absent at 
tree level if all leptonic mixing observable in neutrino oscillations
originated in neutrino (rather than charged lepton) mixing, i.e., 
$ V^e_L = V^e_R=\openone$. For models in which the first two families
have the same couplings, these conditions could be relaxed so that
$V_{L,R}^{d,e}$ mix the first two families only.

Much stricter bounds on these and similar models, including models
with alternative assumptions concerning the fermion mixings, will be
available once rare top decays have been studied at the Tevatron, LHC,
and a future $e^+e^-$ collider, and more stringent bounds on bottom
and tau decays become available {}from existing $b$-factories and
planned charm-$\t$-factories. The rare top decays in particular would
constrain the possibility that quark mixing is restricted to the
$u-c-t$ sector. Improvements in the sensitivity of searches for rare
$K_L$ and $\mu$ decays and $\mu$-$e$ conversion are also highly
desirable.

\section*{Acknowledgements}
We would like to thank J.~Erler and G.~Hiller for helpful discussions
and suggestions.
This work was supported in part by the U.S.\ Department of Energy
Grant No.~EY-76-02-3071 and in part by the Feodor Lynen Program of the
Alexander von Humboldt Foundation (M.P.).

\clearpage

\begin{appendix}

\section{Matrix Elements for $K\to\p\p$\protect\label{KTOPI}}
The evaluation of the hadronic matrix elements in the decay amplitudes
$A_0$ and $A_2$ is clearly a non-perturbative problem. Several methods
have been advocated, and it turns out that chiral perturbation theory
in the large $N_c$ limit, $N_c$ being the number of colors, offers the
best description of $K\to\p\p$ amplitudes \cite{1overN1,1overN2}. 

We will denote the hadronic matrix elements of the operators by
\beq
  \VEV{Q_{qq}^{sd}}_I=
  \langle (2\p)_I| Q_{qq}^{sd} |K^0\rangle\;,
\eeq
where $I=0,2$ denotes the isospin of the two-pion state.
Since only the pseudoscalar part of the effective Lagrangian
contributes to the $K\to\p\p$ amplitudes, the matrix elements of
$\wt{Q}_{qq}^{sd}$ are given by those of $Q_{qq}^{sd}$, with an
additional minus sign. Analogous formulae hold for the matrix elements
of $O_{qq}^{sd}$ and $\wt{O}_{qq}^{sd}$. In the limit corresponding to
the vacuum insertion approximation, chiral perturbation theory the
yields the following matrix elements:
\beqa
  \VEV{Q_{uu}^{sd}}_0={1\over3}\left({2\over N_c}-1\right)X
  \;,&\qquad\qquad&
  \VEV{Q_{uu}^{sd}}_2={\sqrt{2}\over3}\left(1+{1\over N_c}\right)X
  \;,\\[1ex]
  \VEV{Q_{dd}^{sd}}_0={1\over3}\left(1+{1\over N_c}\right)X
  \;,&\qquad\qquad&
  \VEV{Q_{dd}^{sd}}_2=-{\sqrt{2}\over3}\left(1+{1\over N_c}\right)X
  \;,\\[1ex]
  \VEV{O_{uu}^{sd}}_0={1\over3}X-{1\over3N_c}Y
  \;,&\qquad\qquad&
  \VEV{O_{uu}^{sd}}_2=-{\sqrt{2}\over3}X-{1\over3\sqrt{2}N_c}Y
  \;,\\[1ex]
  \VEV{O_{dd}^{sd}}_0=-{1\over3}X
    +{1\over3N_c}\left(3{F_K\over F_{\p}}-2\right)Y
  \;,&\qquad\qquad&
  \VEV{O_{dd}^{sd}}_2={\sqrt{2}\over3}X+{1\over3\sqrt{2}N_c}Y
  \;,
\eeqa
where
\beq
  X=\sqrt{3\over2}F_{\p}\left(m_K^2-m_{\p}^2\right)
    \left(1+{m_{\p}^2\over\L^2_{\chi}}\right)\;,\qquad
  Y=-\sqrt{3\over2}F_{\p}\left({m_K^2\over m_s}\right)^2
    \left(1+{m_{\p}^2\over\L^2_{\chi}}\right)\;,
\eeq
and $\L_{\chi}$ is a parameter in the Lagrangian of chiral perturbation
theory related to the ratio of the $\p$ and $K$ decay constants:
\beq
  {\left(m_K^2-m_{\p}^2\right)\over\L_{\chi}^2}={F_K\over F_{\p}}-1\;.
\eeq

\end{appendix}


\clearpage

\begin{thebibliography}{99}

\bibitem{susyreview} 
For reviews, see M.~Cveti\v{c} and P.~Langacker, in
{\it Perspectives in Supersymmetry}, G.~L.~Kane, ed.\ (World 
Scientific, Singapore, 1998), p.~312, {\tt hep-ph/9707451}; 
P.~Langacker, in {\it Particles, Strings, and Cosmology (PASCOS 98)},
ed.\ P.~Nath
(World Scientific, Singapore, 1999), p.~587, {\tt hep-ph/9805486}.

\bibitem{xdimensions}  See, for example,
M.~Masip and A.~Pomarol, \pr{60}{99}{096005}.

\bibitem{TeVscale}  M.~Cveti\v{c} and P.~Langacker, \pr{54}{96}{3570} and
Mod.~Phys.~Lett.~{\bf A 11}, 1247 (1996);
M.~Cveti\v{c}, D.~A.~Demir, J.~R.~Espinosa,
  L.~Everett and P.~Langacker, \pr{56}{97}{2861}, 
  \pr{58}{98}{119905} (E).

\bibitem{intscale}  See~\cite{TeVscale} and 
G.~Cleaver, M.~Cveti\v{c}, J.R.~Espinosa, L.~Everett and P.~Langacker,
Phys.\ Rev.\ {\bf D57}, 2701 (1998).

\bibitem{muprob} D.~Suematsu and Y.~Yamagishi,
\ijmp{10}{95}{4521}.

\bibitem{gaugemed}  
P.~Langacker, N.~Polonsky and J.~Wang,  Phys. Rev. {\bf D60},
115005 (1999).

\bibitem{cdm}
H.~Cheng, B.~A.~Dobrescu and K.~T.~Matchev,
Phys.~Lett.~{\bf B 439}, 301 (1998); Nucl.~Phys.~{\bf B 543}, 47
(1999).

\bibitem{CDF} F.~Abe et al.\ (CDF collaboration), \prl{79}{97}{2192}.

\bibitem{precision}  J.~Erler and P.~Langacker, \pl{456}{99}{68};
G.C.~Cho et al., \np{531}{98}{65}, \np{555}{99}{651} (E); and
references therein.

\bibitem{indications}  J.~Erler and P.~Langacker, 
Phys.\ Rev.\ Lett.\  {\bf 84}, 212 (2000).
Implications of the atomic parity data alone have recently been discussed
in~\cite{APV}.

\bibitem{APV} R.~Casalbuoni et al.,
Phys.\ Lett.\ {\bf B 460}, 135 (1999); J.L.~Rosner,
Phys.\ Rev.\ {\bf D 61}, 016006 (2000). Earlier references are given
in~\cite{indications}.

\bibitem{diagnostics} For reviews see
M.~Cveti\v{c} and S.~Godfrey, in Proceedings of  {\it Electroweak Symmetry 
Breaking
and Beyond the Standard Model}, eds.\ T.~Barklow, S.~Dawson, H.~Haber 
and J.~Siegrist (World Scientific, Singapore, 1995); 
A.~Leike, \prep{317}{99}{143}.

\bibitem{altspec} L.~Everett, P.~Langacker,
M.~Pl\"umacher, and J.~Wang, {\tt hep-ph/0001073}.

\bibitem{nonuniversal} Studies that do include nonuniversality and
 flavor changing effects include: 
T.~K.~Kuo and N.~Nakagawa, \pr{31}{85}{1161}; \pr{32}{85}{306};
K.~K.~Gan, \pl{209}{88}{95};
E.~Nardi, Phys.\ Rev.\  {\bf D48}, 
1240 (1993); talk presented at {\it Particles \& Fields 92: 7th Meeting 
of the Division of Particles Fields of the APS (DPF 92)}, Batavia, IL, 
10-14 November 1992, {\tt hep-ph/9211246}; 
B.~Holdom, Phys.\ Lett.\  {\bf B339}, 114 (1994); 
X.~Zhang and B.-L.~Young, \pr{51}{95}{6584}; 
B.~Holdom and M.~V.~Ramana, Phys.\ Lett.\  {\bf B365}, 309 (1996).

\bibitem{CHL}
S.~Chaudhuri, S.W.~Chung, G.~Hockney, and J.~Lykken,
\np{456}{95}{89}.

\bibitem{CHL5}
G.~Cleaver, M.~Cveti\v{c}, J.R.~Espinosa, L.~Everett, 
and P.~Langacker, \np{525}{98}{3}; 
G.~Cleaver, M.~Cveti\v{c}, J.R.~Espinosa, L.~Everett, P.~Langacker,
and J.~Wang, \prd{59}{055005}(1999), ibid. 115003 (1999).

\bibitem{pdg}
 C.~Caso et al., Eur.~Phys.~J.~{\bf C3}, 1 (1998) 
    and 1999 off-year partial update for the 2000 edition available on 
    the PDG WWW pages (URL: {\tt http://pdg.lbl.gov/}). 

\bibitem{exotic}  
P.~Langacker and D.~London, Phys.~Rev.~{\bf D 38}, 886 (1988); 
E.~Nardi, E.~Roulet and D.~Tommasini, Phys.~Rev.~{\bf D 46}, 3040 (1992)
and Nucl.~Phys. {\bf B386}, 239 (1992).

\bibitem{Zprime}
L. S. Durkin and P. Langacker, \pl{166}{86}{436};
P.~Langacker and M.~Luo, \pr{45}{92}{278}.

\bibitem{kinmix}
N.~S.~Kroll, T.~D.~Lee, and B.~Zumino, Phys.~Rev.~{\bf 157}, 1376 (1967);
B. Holdom, Phys.~Lett.~{\bf B166}, 196 (1986);
F.~Del Aguila, M.~Cveti\v{c} and P.~Langacker, Phys.~Rev.~{\bf D52},
37 (1995); 
F.~del Aguila, M.~Masip, M.~Perez-Victoria, Acta Phys.~Polon.~{\bf 
 B27}, 1469 (1996), {\tt hep-ph/9603347};
K.~S.~Babu, C.~Kolda, and J.~March-Russell, \pr{54}{96}{4635}.

\bibitem{okada}
Y.~Kuno and Y.~Okada, {\tt hep-ph/9909265}.

\bibitem{tau}
S.~Gentile and M.~Pohl, Phys.~Rep.~{\bf 274}, 287 (1996).

\bibitem{gan}
K.~K.~Gan, in \cite{nonuniversal}.

\bibitem{sindrum}
P.~Wintz, in {\it Proceedings of the First Int.\ Symp.\ on Lepton and
Baryon Number Violation}, Trento, 1998, (IoP Publishing, Bristol and
Philadelphia, 1999), eds.\ H.~V.~Klapdor-Kleingrothaus and 
I.~V.~Krivosheina, p.~534.

\bibitem{bernabeu}
J.~Bernab\'eu, E.~Nardi and D.~Tommasini, 
Nucl.~Phys.~{\bf B 409}, 69 (1993).

\bibitem{buras_rev}
G.~Buchalla, A.~J.~Buras and M.~E.~Lautenbacher, 
  Rev.~Mod.~Phys.~{\bf 68}, 1125 (1996).

\bibitem{Kdecays}
J.~L.~Ritchie and S.~G.~Wojcicki, Rev.~Mod.~Phys.~{\bf 65}, 1149 (1993).

\bibitem{Ktopiee}
J.~F.~Donoghue and F.~Gabbiani, \pr{51}{95}{2187}.

\bibitem{KTEV}
J.~Whitmore, preprint {\tt FERMILAB-CONF-99-266-E},
{\it To be published in the proceedings of 1999 Chicago 
Conference on Kaon Physics (K 99), Chicago, IL, 21-26 Jun 1999}.

\bibitem{DM_matr_elem}
J.-M.~G\'erard, W.~Grimus, A.~Raychaudhuri and G.~Zoupanos, 
\pl{140}{84}{349};\\
F.~Gabbiani, E.~Gabrielli, A.~Masiero and 
L.~Silvestrini, \np{477}{96}{321}.

\bibitem{bnl}
D.~Ambrose et al.\ (BNL E871 collaboration), Phys.~Rev.~Lett.~{\bf
  81}, 5734 (1998).

\bibitem{buchalla}
G.~Buchalla, Presented at {\it International Europhysics Conference 
on High-Energy Physics} (EPS-HEP 99), Tampere, Finland, 15-21 Jul 1999,
{\tt hep-ph/9912369}.

\bibitem{cleo}
R.~Godang et al.\ (CLEO collaboration), report {\tt CLEO CONF 98-22},
submitted to the {\it XXIXth Int.\ Conf.\ on High Energy
Physics (ICHEP 98)}, Vancouver, British Columbia, Canada, 
23-29 Jul.\ 1998.

\bibitem{topdecays}
T.~Han, R.D.~Peccei, and X.~Zhang, \np{454}{95}{527};\\
T.~Han, K.~Whisnant, B.-L.~Young, and X.~Zhang, \pr{55}{97}{7241};\\
J.~L.~D\'{\i}az-Cruz, M.A.~P\'erez, G.~Tavares-Velasco, and J.J.~Toscano,
Phys.~Rev.~{\bf D 60}, 115014 (1999).

\bibitem{topcharm}
T.~Han, M.~Hosch, K.~Whisnant, B.~Young and X.~Zhang,
Phys.\ Rev.\  {\bf D58}, 073008 (1998);\\
F.~del Aguila, J.~A.~Aguilar-Saavedra and R.~Miquel,
\prl{82}{99}{1628};\\
T.~Han and J.~Hewett, Phys.~Rev.~{\bf D 60}, 074015 (1999);\\
F.~del Aguila, J.~A.~Aguilar-Saavedra and Ll.~Ametller, 
Phys.~Lett.~{\bf B 462}, 310 (1999);\\
F.~del Aguila and J.~A.~Aguilar-Saavedra, {\tt hep-ph/9909222}.

\bibitem{Bsmixing}
S.~Willocq, presented at the
{\it 2000 Aspen Winter Conference on Particle Physics},  January 16 - 22, 2000, 
Aspen, CO.

\bibitem{MEGA}
M.~L.~Brooks et al.\ (MEGA Collaboration),
Phys.\ Rev.\ Lett.\  {\bf 83}, 1521 (1999).

\bibitem{atmo}
Y.~Fukuda {\it et al.}  (Super-Kamiokande Collaboration),
Phys.\ Rev.\ Lett.\  {\bf 81}, 1562 (1998).

\bibitem{E6nardi}
A.~Masiero, D.~V.~Nanopoulos and A.~I.~Sanda, \prl{57}{86}{663};\\
G.~C.~Branco and C.~Q.~Geng, \prl{58}{87}{969};\\
E.~Nardi, Phys.~Rev.~{\bf D 48}, 3277 (1993); 
  Phys.~Rev.~{\bf D 49}, 4394 (1994);\\
E.~Nardi and T.~Rizzo, Phys.~Rev.~{\bf D 50}, 203 (1994).

\bibitem{1overN1}
R.~S.~Chivukula, J.~M.~Flynn and H.~Georgi, \pl{171}{86}{453}.

\bibitem{1overN2}
A.~J.~Buras and J.-M.~G\'erard, \np{264}{86}{371};\\
W.~A.~Bardeen, A.~J.~Buras and J.-M.~G\'erard, \np{293}{87}{787};\\
G.~Buchalla, A.~J.~Buras and M.~K.~Harlander, \np{337}{90}{313}.


\end{thebibliography}
\end{document}